\documentclass[preprint]{sigplanconf}

\usepackage{amsmath}
\usepackage{enumitem}
\usepackage{tikz}

\usetikzlibrary{matrix}

\newcommand{\Comment}[1]{}

\newcommand{\etal}{\textit{et~al.}}
\newcommand{\eg}{\textit{e.g.}}

\newcommand{\ie}{\textit{i.e.}}
\newcommand{\FUNCTORMIN}{\texttt{FUNCTOR\_MIN}}
\newcommand{\FUNCTOR}{\texttt{FUNCTOR}}
\newcommand{\APPLICATIVEMIN}{\texttt{APPLICATIVE\_MIN}}
\newcommand{\APPLICATIVE}{\texttt{APPLICATIVE}}
\newcommand{\ALTERNATIVEMIN}{\texttt{ALTER\_MIN}}
\newcommand{\ALTERNATIVE}{\texttt{ALTER}}
\newcommand{\MONADZPLUS}{\texttt{MONAD0P}}
\newcommand{\MONADZPLUSMIN}{\texttt{MONAD0P\_MIN}}
\newcommand{\MONADMIN}{\texttt{MONAD\_MIN}}
\newcommand{\MONAD}{\texttt{MONAD}}
\newcommand{\mkFUNCTOR}{\texttt{mk\_Functor}}
\newcommand{\mkAPPLICATIVE}{\texttt{mk\_Applicative}}
\newcommand{\mkMONAD}{\texttt{mk\_Monad}}
\newcommand{\ApplicativeMinToFunctorMin}{\texttt{App\_Min\_To\_Fun\_Min}}
\newcommand{\MonadMinToApplicativeMin}{\texttt{Mona\_Min\_To\_App\_Min}}
\newcommand{\functor}{\texttt{functor}}
\newcommand{\applicative}{\texttt{applicative}}
\newcommand{\monad}{\texttt{monad}}
\newcommand{\signature}{\texttt{signature}}

\begin{document}

\setlength{\pdfpageheight}{\paperheight}
\setlength{\pdfpagewidth}{\paperwidth}




\title{Close Encounters of the Higher Kind}
\subtitle{Emulating Constructor Classes in Standard ML}

\authorinfo{Yutaka Nagashima}
           {Data61, CSIRO / NICTA}
           {yutaka.nagashima@nicta.com.au}
\authorinfo{Liam O'Connor}
           {UNSW Australia and Data61, CSIRO / NICTA}
           {liamoc@cse.unsw.edu.au}

\maketitle

\emph{We implement a library for encoding \emph{constructor classes} in Standard ML, including
elaboration from minimal definitions, and automatic instantiation of superclasses.}




\section{Introduction}
In our recent work \cite{nagashima16} on automating Isabelle proofs, we discovered that
several proof search problems can be elegantly expressed as a monadic program. 
Unfortunately, Standard ML does not natively support the kinds of polymorphism required
to easily express a \texttt{Monad} abstraction, nor similar abstractions such as \texttt{Applicative} and \texttt{Functor}\footnote{Not to be confused with an ML \functor{}}. In this paper, we present a technique
for encoding constructor classes such as \texttt{Monad}, which relies only on the Standard ML module system.

Several others have attempted to enable constructor classes in Standard ML by changing the language. While it is tempting to customise the language by adding new features,
new features tend to cause duplication \citep{dreyer07}
and inconsistency \citep{kuncar15}.
Furthermore, avoiding language extensions makes our approach transferable to all other ML dialects with a module system.

Our contributions are twofold: we develop a usable library for monads, monad transformers, applicatives, and more in Standard ML, and demonstrate an elegant technique using ML functors to elaborate minimal definitions of each abstraction to avoid code duplication. For example, given a minimal definition of the \texttt{list} monad, \eg{} \texttt{return} and \texttt{bind},
our library derives other basic functions, such as \texttt{>=>}, \texttt{join}, 
and \texttt{liftM} automatically.
Moreover, using the hierarchical relationship among constructor classes,
our library automatically instantiates \texttt{list} as a member of the parent classes,
\eg{} \texttt{applicative} and \texttt{functor}.
Thus, for each \texttt{monad}, users can derive more than twenty functions 
from two manually written functions, \ie{} \texttt{return} and \texttt{bind}.

\section{Constructor Classes in Standard ML}
\begin{figure}
\begin{tikzpicture}
  \matrix (m) [matrix of math nodes,row sep=2em,column sep=2em,minimum width=2em]
  {
     \FUNCTORMIN{} & \APPLICATIVEMIN{} & \MONADMIN{}\\
     \FUNCTOR{}    & \APPLICATIVE      & \MONAD{}   \\};
  \path[dashed,->]
    (m-1-1) edge [bend left] node [label={[label distance=0em]0:{\mkFUNCTOR}}]     {} (m-2-1)
    (m-1-2) edge [bend left] node [label={[label distance=0em]0:{\mkAPPLICATIVE}}] {} (m-2-2)
    (m-1-3) edge [bend left] node [label={[label distance=0em]0:{\mkMONAD}}]       {} (m-2-3)
    (m-1-2) edge [bend right] node [label={[label distance=-0.5em]90:{\ApplicativeMinToFunctorMin \quad\quad}}] {} (m-1-1)
    (m-1-3) edge [bend right] node [label={[label distance=-0.5em]90:{\quad\quad\quad\MonadMinToApplicativeMin}}] {} (m-1-2);
  \path[->]
    (m-2-1) edge   (m-1-1)
    (m-2-2) edge   (m-1-2)
    (m-2-3) edge   (m-1-3)
    (m-2-2) edge   (m-2-1)
    (m-2-3) edge   (m-2-2);
\end{tikzpicture}
\caption{Automatic instantiation and function derivation.}
\label{fig:derivation}
\end{figure}
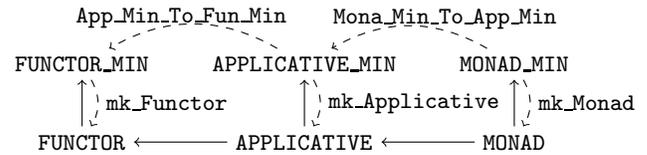

Figure \ref{fig:derivation} shows the structure of the class hierarchy as it is implemented
in our library. Each node represents a ML \signature{}.
Straight arrows stand for subtyping relations, 
whereas dashed arrows with labels stand for ML \functor{}s and their names.
The ML \texttt{functor}s expressed as vertical dashed arrows, 
\eg{} \mkMONAD{}, produce full definitions
of constructor classes from the corresponding minimal definitions;
those expressed as horizontal dashed arrows, \eg{} \MonadMinToApplicativeMin, 
generalise minimal definitions for a class to its superclass.

\setlength{\textfloatsep}{10pt}
For example, the following code snippets show the specifications of \MONADMIN{} and \MONAD{}.
\vspace*{1em}\\
\noindent
\texttt{signature MONAD\_MIN =}\\
\texttt{sig}\\
\hspace*{0.5em}type 'a monad;\\
\texttt{\hspace*{0.5em}val return:'a -> 'a monad;}\\
\texttt{\hspace*{0.5em}val bind:'a monad -> ('a -> 'b monad) -> 'b monad;\\
end;}\\
\noindent
\texttt{
signature MONAD =\\
sig\\
\hspace*{0.5em}include APPLICATIVE MONAD\_MIN;\\
\hspace*{0.5em}sharing type monad = applicative;\\
\hspace*{0.5em}val liftM : ('a -> 'b) -> ('a monad -> 'b monad);\\
\hspace*{0.5em}val join: \dots; val forever: \dots; val \dots\\
end;}
\vspace*{1em}\\
\noindent
Since every \monad{} is \applicative{},
we express this subtyping relation
using the ML keyword \texttt{include}.
In Haskell, this relation is expressed as 
\texttt{class Applicative m => Monad m} .

In order to create a concrete instance of a constructor class, the user
merely supplies its minimal definition. For example, one can instantiate 
the type constructor \texttt{list}
as a member of \MONAD{} by defining the following module.
\vspace*{1em}\\
\noindent
\texttt{structure ListMonadMin:MONAD\_MIN =\\
struct\\
\hspace*{0.5em}type 'a monad = 'a list;\\
\hspace*{0.5em}fun return x      = [x];\\
\hspace*{0.5em}fun bind seq func = List.concat (map func seq);\\
end;
}
\vspace*{1em}\\
\noindent
Then, passing \texttt{ListMonadMin} to the ML \functor{} \mkMONAD{}
produces a full-fledged instance of \MONAD{}:
\vspace*{1em}\\
\noindent
\texttt{structure ListMonad:MONAD = mk\_Monad(ListMonadMin)}
\vspace*{1em}\\
\noindent
A minimal instance of \APPLICATIVE{} can also be produced from our structure by using the appropriate functors:
\begin{verbatim}
structure ListAppMin:APPLICATIVE_MIN 
   = Mona_Min_To_App_Min(ListMonadMin) 
\end{verbatim}
\noindent
Note that these ML \functor{}s go in the same direction as the subtyping relation, unlike
the elaboration functors such as \mkMONAD{}.  The following shows the definition of the
\MonadMinToApplicativeMin{} functor.
\vspace*{1em}\\
\noindent
\texttt{functor \MonadMinToApplicativeMin{} (Min:MONAD\_MIN) =\\
struct\\
\hspace*{0.5em}open Min;\\
\hspace*{0.5em}type 'a applicative = 'a monad;\\
\hspace*{0.5em}val pure = return;\\
\hspace*{0.5em}fun <*> (fs, xs) = bind fs (fn fs' => \\
\hspace*{10em}bind xs (fn xs' => \\
\hspace*{10em}return (fs' xs')));\\
end : APPLICATIVE\_MIN ;}
\vspace*{1em}\\
\noindent
The ML \functor{} \MonadMinToApplicativeMin{} produces instances of \APPLICATIVEMIN{} in terms of \MONADMIN{} functions.
This is in contrast with the constructor classes in Haskell 
where \texttt{return} is defined as \texttt{pure}.
It is this inversion that enables our library to 
derive superclass instances for a given type constructor.

The elaboration functor \mkMONAD{} is defined as follows.
\vspace*{1em}\\
\noindent
\texttt{
functor mk\_Monad (Min : MONAD\_MIN): MONAD =\\
struct\\
\hspace*{0.5em}type 'a monad = 'a Min.monad;\\
\hspace*{0.5em}structure App\_Min = Mona\_Min\_to\_App\_Min (Min);\\
\hspace*{0.5em}structure App = mk\_Applicative (App\_Min);\\
\hspace*{0.5em}open App Min;\\
\hspace*{0.5em}fun liftM f m = bind m (fn m' => return (f m'));\\
\hspace*{0.5em}fun join n    = \ldots; fun forever a = \dots; fun \ldots\\
end;
}
\vspace*{1em}\\
\noindent
Apart from producing the various \MONAD{} functions,
\texttt{mk\_Monad} instantiates \texttt{list} as a member of \texttt{\APPLICATIVE{}}
by elaborating the result of the functor \MonadMinToApplicativeMin{} with \mkAPPLICATIVE{},
which in turn instantiates the \FUNCTOR{} class similarly.

\begin{figure}
\begin{tikzpicture}
  \matrix (m) [matrix of math nodes, row sep=0.6em, column sep=0.1em]{
    & \ALTERNATIVEMIN{} & & \MONADZPLUSMIN{} \\
      \APPLICATIVEMIN{} & & \MONADMIN{}& \\
    & \ALTERNATIVE{}    & & \MONADZPLUS{} \\
      \APPLICATIVE{}    & & \MONAD{} & \\};
  \path[-stealth]
    (m-3-2) edge (m-4-1) 
    (m-4-1) edge (m-2-1) 
    (m-3-4) edge (m-3-2) 
            edge (m-4-3) 
            edge (m-1-4) 
    (m-3-2) edge (m-1-2) 
    (m-4-3) edge (m-4-1) 
            edge (m-2-3) 
            ;
  \path[dashed,->]
    (m-1-2) edge[bend right] (m-2-1)
    (m-1-4) edge[bend right=15] (m-1-2)
            edge[bend right] (m-2-3)
    (m-2-3) edge[bend right=15] (m-2-1) 
    ;
  \path[dashed,->]
    (m-1-2) edge[bend right] (m-3-2)
    (m-1-4) edge[bend right] (m-3-4)
    (m-2-3) edge[bend right] (m-4-3)
    (m-2-1) edge[bend right] (m-4-1)
    ;
\end{tikzpicture}
\caption{Diamond case.}
\label{fig:diamond}
\end{figure}
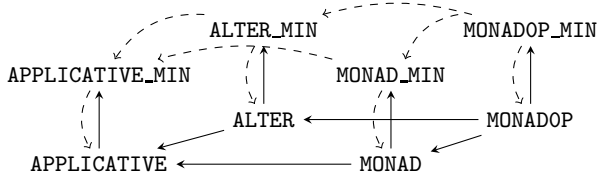

We formalise monad transformers as ML \texttt{functor}s, too.
For instance, the state monad transformer is a \texttt{functor} 
that takes two modules, the minimal definition of the base monad and a module containing just the type of the state, and produces a minimal definition of the transformed monad.
\Comment{I developed a special signature for the transformed monads. Should I mention this?
It is similar to \MONADMIN{} but embellished with the base monad and the function \texttt{lift}.}

\setlength{\textfloatsep}{10pt}
\section{Corner Cases \Comment{Better name for this section?}}
Some functions in Haskell involve multiple classes, such as \texttt{foldM}:
\begin{verbatim}
foldM :: (Foldable t, Monad m) 
      => (b -> a -> m b) -> b -> t a -> m b
\end{verbatim}
We formalise these as ML \texttt{functor}s
that take multiple modules conforming to the appropriate signatures and return a module containing the function.

We can easily extend our approach to other constructor classes,
even if they involve multiple inheritance.
Figure \ref{fig:diamond} shows an example of such a case.
Since our library is based on statically known mathematical properties,
we avoid so-called \emph{diamond problem}s.
For instance, given a type constructor of \MONADZPLUS{} in Figure \ref{fig:diamond},
it does not matter semantically
from which of \ALTERNATIVE{} and \MONAD{} 
this type constructor inherits the methods of \APPLICATIVE{},
as both of them have the same properties.

\section{Comparison and Related Work}
Our approach offers some benefits over traditional Haskell type classes. In particular, the ML module system allows more flexibility, as more than one instance can be provided for a given type. This flexibility is appreciated in constructor classes, too --- for example, there are two perfectly valid \texttt{Applicative} instances for lists, one with a cartesian and one with a pairwise product operation. In Haskell, this necessitates the use of the \texttt{newtype} feature for one of the instances. In ML, both instances are equally natural.

\citet{wehr08} first introduced an approach to translate 
Haskell type classes in ML modules. 
They discussed that their scheme is not able to handle constructor classes, 
nor translate either recursive class constraints or default definitions into ML modules,
while we addressed all of these. One example of a recursive class constraint would be:
\begin{verbatim}
instance (Monad f, Monad g) => Monad (f :*: g)
\end{verbatim}
We express these using ML \texttt{functor}s:
in this case, we define a \texttt{functor} \texttt{mk\_ConsProd},
which takes two modules of \MONADMIN{} and returns a module of \MONADMIN{}.
Even though we can define \texttt{mk\_ConsProd} parametrically, 
two concrete type constructors \texttt{f} and \texttt{g} must be supplied in order to instantiate \MONADMIN{} for \texttt{f :*: g}.
\Comment{Note that both \texttt{f} and \texttt{g} are type constructors.
Therefore, we defined \texttt{(:*:)} itself using the ML module system.}

\Comment{Our library does not support recursive class constraints if they are used in
class declarations instead of instantiation.
But I have not seen such a case.}

Our approach is similar to the library code in \citet{dreyer07}; however, we additionally support constructor classes, instance elaboration, and automatic instantiation of superclasses.
We did not, however, extend the language as they did, as we did not wish to deviate from Standard ML, although we foresee no fundamental problems incorporating  their implicit typing scheme into our library.
Furthermore, we chose to express class hierarchies with flat module structures,
while they did so hierarchically.
Our choice allows users to avoid nested qualifiers, 
\eg{} \texttt{ListMonad.Applicative.Fmap.<\$},
resulting in less verbose code in the absence of any implicit typing mechanism.

\citet{scott} seems to have employed a similar approach to ours, 
but in OCaml, suggesting that our technique is transferable to other ML dialects.
There are also attempts to model type and constructor classes using 
features from the imperative object oriented programming paradigm.
We purposefully avoided these deviations from Standard ML.
\Comment{Should I cite theses? "Modular Implicits" by Leo White and "Type Classes as Objects and Implicits".}
\Comment{Imperative OOP is a right expression? 
I do not have problems with subtyping relations. 
In fact, I am using subtyping relations in my framework.
What I want to avoid is dynamic typing.
I guess "Type Class as Objects and Implicits" is something I want to avoid;
the paper talks about "retroactive extension",
which does not sound like giving a static guarantee about programs.}

\section{Current Status and Future Work}
We previously developed \citep{nagashima16} a proof automation tool for Isabelle using this library, and our experience with it was positive.
\Comment{I should cite my draft. But I have to submit the paper to arXiv. first.}
However, every library has room for improvement.
We are working to include other constructor classes such as \texttt{Arrow} into this framework.
In our approach, our \texttt{MONAD} module could also generate an instance  of \texttt{ARROW},
once again eliminating the Haskell use of \texttt{newtype} for \texttt{Kleisli} arrows.

Furthermore, we plan to support multiple minimal definitions to instantiate some constructor classes.
For example, we presented a minimal definition of \MONAD{} with \texttt{return} and \texttt{bind} above,
but we could provide a minimal definition of \MONAD{} 
with \texttt{return}, \texttt{fmap}, and \texttt{join} instead.
It is up to the user's preference which minimal definition is easier to write.
Since they are equivalent, 
we can write a \texttt{functor} that derives one from the other, 
providing multiple options to users.
\Comment{This is fairly easy; I can finish this and add this to Corner Cases.}\\
\Comment{I have not talked about performance issues.
I reckon that derived functions are not very efficient;
however, users can provide their efficient implementation and overload the module easily.
I.e. 
structure FastListMonad : MONAD = 
struct 
  open ListMonad; 
  fun $<*>$ (fs, xs) = fast\_implementation;
end;}

\acks
We thank Gabriele Keller for stimulating discussions on ways of representing constructor classes with modules.
NICTA is funded by the Australian Government through the Department of Communications and the Australian Research Council through the ICT Centre of Excellence Program.


\bibliographystyle{abbrvnat}


\end{document}